\documentclass{ws-ijmpc}
\usepackage{amssymb}
\usepackage{graphicx}
\usepackage{dsfont}

\begin{document}

\markboth{Frank Emmert-Streib}
{A Heterosynaptic Learning Rule for Neural Networks}

\catchline{}{}{}{}{}
%%%%%%%%%%%%%%%%%%%%%%%%%%%%%%%%%%%%%%%%%%%%%%%%%%%%%%%%%%%%%%%%%%%%%

\title{A Heterosynaptic Learning Rule for Neural Networks}

\author{Frank Emmert-Streib\footnote{Present address: Stowers Institute for Medical Research, 1000 E. 50th Street, Kansas City, MO 64110, USA}}
\address{Institut f\"ur Theoretische Physik, Universit\"at Bremen, Otto-Hahn-Allee, 28334 Bremen, Germany\\
fes@stowers-institute.org}

\maketitle

\begin{history}
\received{Day Month Year}
\revised{Day Month Year}
\end{history}

\begin{abstract}

In this article we intoduce a novel stochastic Hebb-like learning rule for neural networks that is neurobiologically motivated. This learning rule combines features of unsupervised (Hebbian) and supervised (reinforcement) learning and is stochastic with respect to the selection of the time points when a synapse is modified. Moreover, the learning rule does not only affect the synapse between pre- and postsynaptic neuron, which is called homosynaptic plasticity, but effects also further remote synapses of the pre- and postsynaptic neuron. This more complex form of synaptic plasticity has recently come under investigations in neurobiology and is called heterosynaptic plasticity. We demonstrate that this learning rule is useful in training neural networks by learning parity functions including the exclusive-or (XOR) mapping in a multilayer feed-forward network. We find, that our stochastic learning rule works well, even in the presence of noise. Importantly, the mean learning time increases with the number of patterns to be learned polynomially, indicating efficient learning.

\end{abstract}

\section{Introduction}

What are the laws that regulate learning on a neuronal level in animals or humans? So far this important question is open, however, the imagination one has for a biological learning rule is that the synaptic weights are changed according to a local rule. In the context of neural networks {\it{local}} means that only the adjacent neurons of a synapse contribute to changes of the synaptic weight. Such a mechanism with respect to synaptic strengthening was proposed by Donald Hebb \cite{h1949} in 1949 and experimentally found by T. Bliss and T. Lomo \cite{blisslomo_1973}. In a biological terminus Hebbian learning is called {\it{long-term potentiation}} (LTP). Experimentally as well as theoretically there is a great body of investigations aiming to formulate precise conditions under which learning in neural networks takes place. For example the influence of the precise timing of pre- and postsynaptic neuron firing \cite{markramluebke_1997,kempter_1999} or the duration of a synaptic change (for a review see \cite{koch_1999}) termed {\it{short}} or {\it{long-term plasticity}} have been studied extensively. All of these contributions share the locality condition proposed by Hebb \cite{h1949}. In this article we present a novel stochastic Hebb-like learning rule inspired by experimental findings about heterosynaptic plasticity \cite{fitzsimonds_1997}. This form of neural plasticity affects not only the synpase between pre- and postsynaptic neuron in which a synaptic modification was induced, but also further remote synapses of the pre- and postsynaptic neuron. Additionally, we demonstrate that this learning rule can be successfully applied to train multilayer neural networks.

This paper is organized as follows. In section \ref{intro_nn} we give the motivation for our learning rule by a summary of experimental observations concerning synaptic plasticity and properties of biological and artificial neural networks as far as they are useful for a better understanding of our learning rule. In section \ref{def_lr} we propose our learning rule and give a mathematical definition. We investigate our learning rule in section \ref{results} by numerical simulations. In section \ref{discussion} we discuss and compare our stochastic learning rule with other learning rules. This article ends in section \ref{end} with a conclusion and an outlook on further investigations.

%%%%%%%%%%%%%%%%%%%%%%%%%%%%%%%%%%%%%%%%%%%%%%%%%%%%%%%%%%%%%%%%%%%%%%%

\section{Overview of biological and artificial learning in neural networks}
\label{intro_nn} 

One property that have all neural networks in common, biological as well as artificial, is that there are two different processes taking place simultaneously. The first process concerns signal processing and the second learning. Signal processing is reflected by the time dependent activity $x_i(t)$ of a neuron $i$, whereas learning concerns the dynamical behavior of the synaptic weights $w_{ij}(t)$ between two neurons $i$ and $j$ in the network. One major difference between both dynamics is that they occur on different timescales. Normally, learning is much slower than the neural activity. Despite our focus in this article on the learning dynamics, we can not neglect a treatment of the neural activity, because both processes are coupled and influence each other.

\begin{figure}[t!]
\centering
\begin{minipage}[c]{0.45\textwidth}
\centerline{\psfig{file=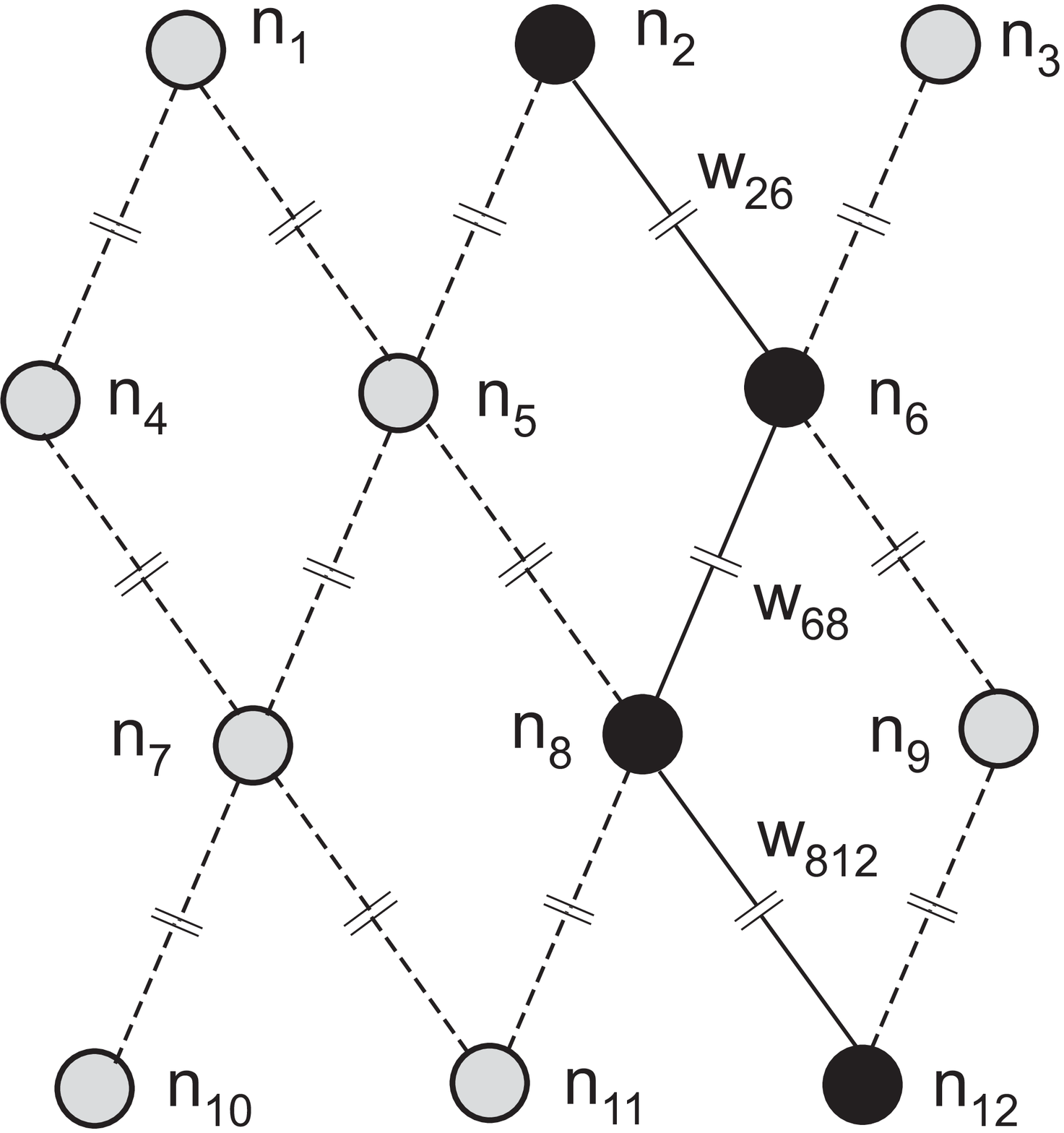,width=50mm}}
\end{minipage}
\centering
\begin{minipage}[c]{0.45\textwidth}
\centerline{\psfig{file=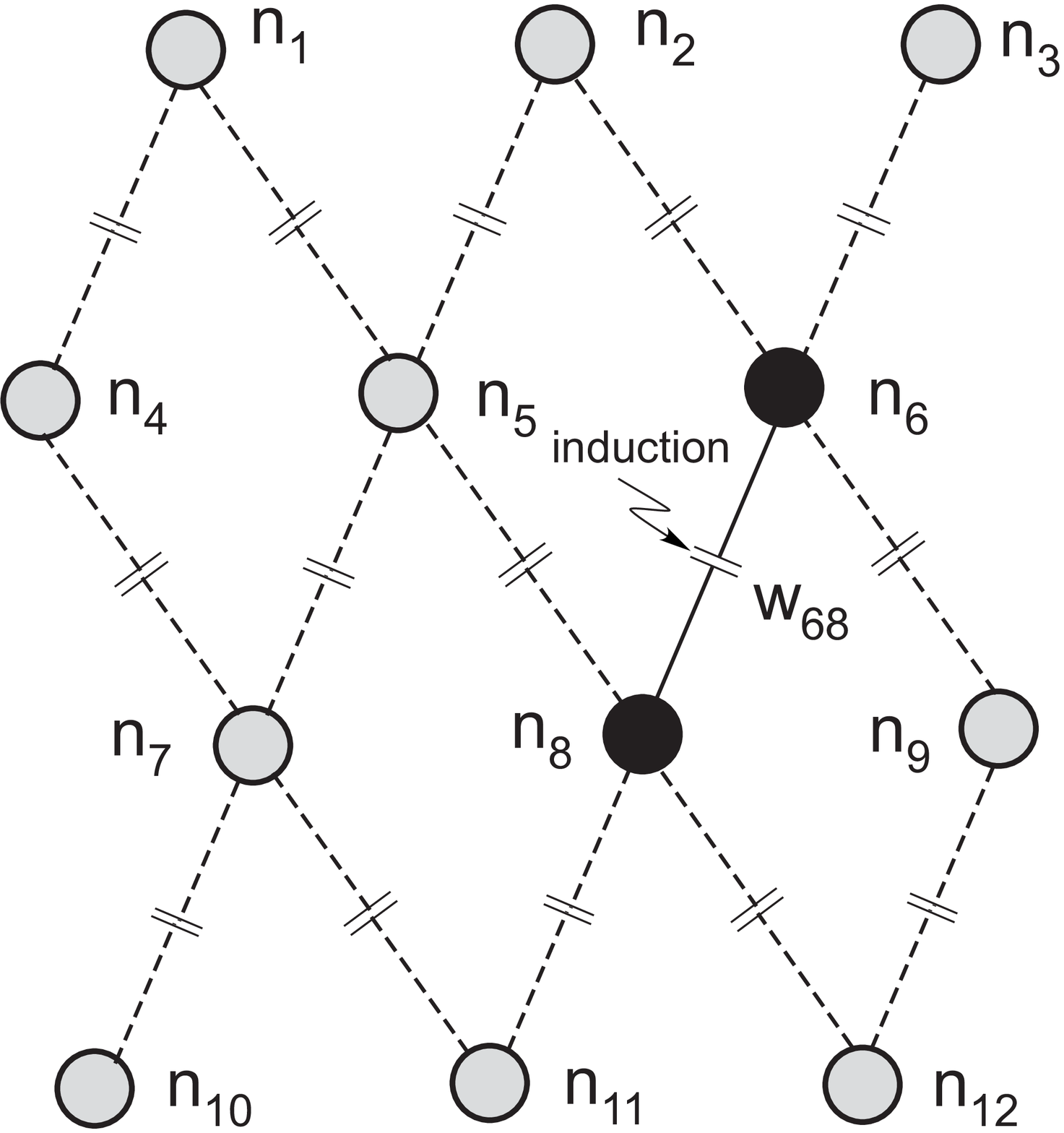,width=50mm}}
\end{minipage}
\vspace*{8pt}
\caption{Schematically depiction of a feed-forward neural network with time direction from top to down. Left: Visualization of the {\it{synaptic tagging}} mechanism experimentally found by Frey et al.. Right: Homosynaptic plasticity induced by the simmultanious activity of neuron $n_6$ and $n_8$ within a certain time window.}\label{fig1}
\end{figure}

Figure \ref{fig1} shows a schematic neural network consisting of $12$ neurons.
The synapses are not drawn directly from neuron to neuron but in two pieces. This shall depict the synaptic cleft of chemical synapses. The reason for this becomes more clear, when we describe our learning rule below. The left figure describes a signal path within a feed-forward network involving the neurons $n_2,n_6,n_8,n_{12}$ and the synapses between these neurons $w_{\mathrm{26}},w_{\mathrm{68}},w_{\mathrm{812}}$. In this and all following figures we suppose that the signal flow and, hence, the orientation of the path, is from the top to the bottom. The neurons (synapses), which were actively involved in this signal processing, are drawn as black circles (full lines). Concerning this information flow, Frey et al. \cite{freymorris_1997} found in the hippocampus of rats in vivo that there is a {\it{synaptic tagging}} mechanism. This mechanism tagges synapses which were repeatly involved in information processing within a certain time window of up to 1.5 hours. If one of these synapses is restimulated within this time interval then a synaptic modification is induced. One can interpret this as a kind of echo or memory within the neural network of past activity. Hence, the left Fig. \ref{fig1} can be interpreted in a way that the depicted path from neuron $n_2$ to $n_{12}$ is not the actual information flow, but the reflection of recent past activity, which the neurons and synapses can remember by an additional degree of freedom.

Suppose now, that this signal flow caused a synaptic modification on $w_{\mathrm{68}}$ as depicted in the right Fig. \ref{fig1}. This situation corresponds to the so called Hebbian learning \cite{h1949}. Necessary conditions for this kind of learning are that the neurons, surrounding the synapse, were both active within a certain time window, which is in the $ms$ range, and that the presynaptic neuron fires before the postsynaptic neuron \cite{markramluebke_1997}. In biological terms Hebbian learning is also called {\it{long-term potentiation}} (LTP), because it strengthens the synaptic weight in contrast to {\it{long-term depression}} (LTD), which weakens the synaptic weight, if the spiking time points of pre- and postsynaptic neuron are reversed. However, both kinds of learning, LTP as well as LTD, have one common ground, they are homosynaptic in respect to the number of synapses which are changed.

\begin{figure}[t!]
\centering
\begin{minipage}[c]{0.45\textwidth}
\includegraphics[width=51mm]{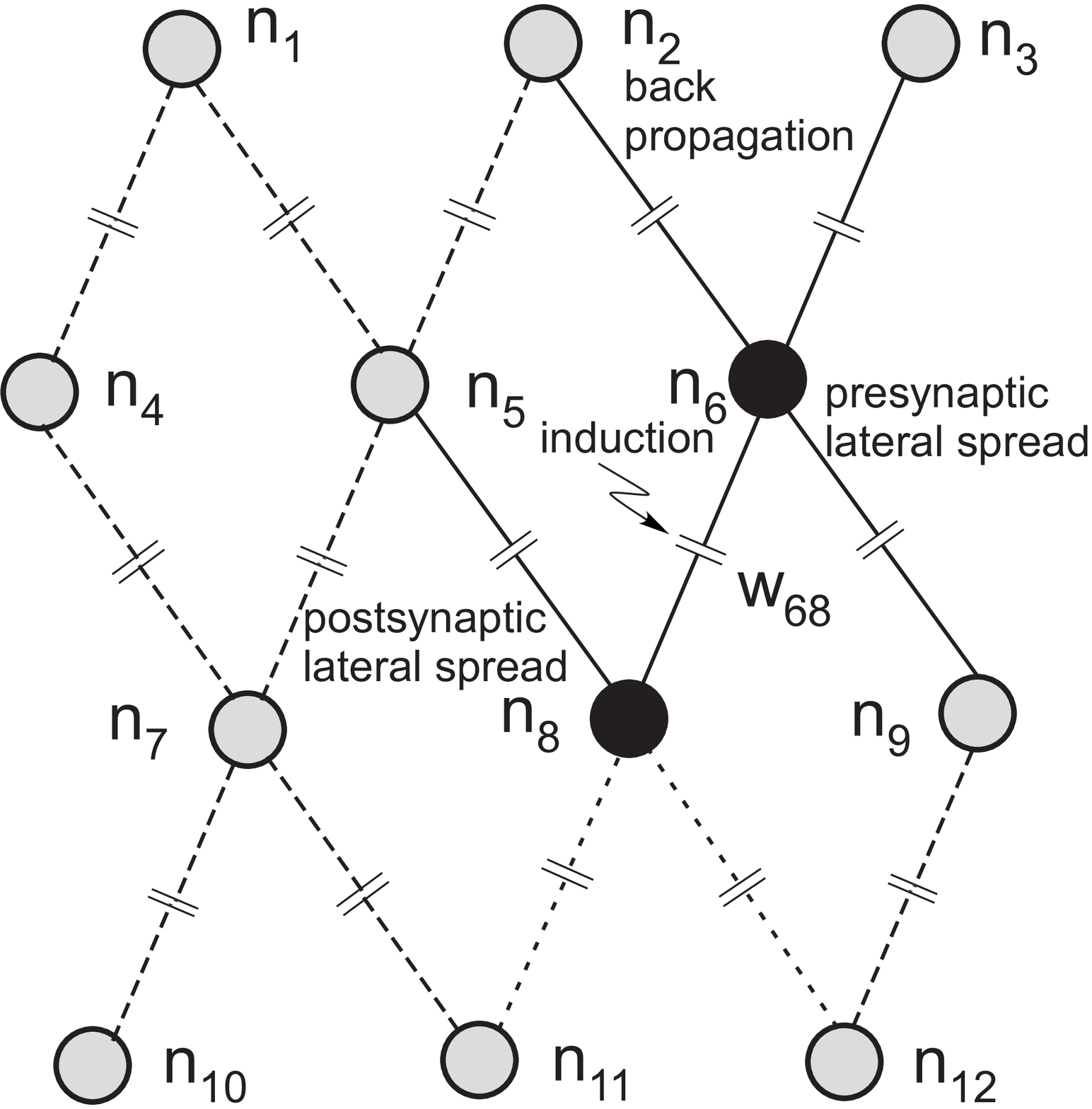}
\end{minipage}
\centering
\begin{minipage}[c]{0.45\textwidth}
\includegraphics[width=50mm]{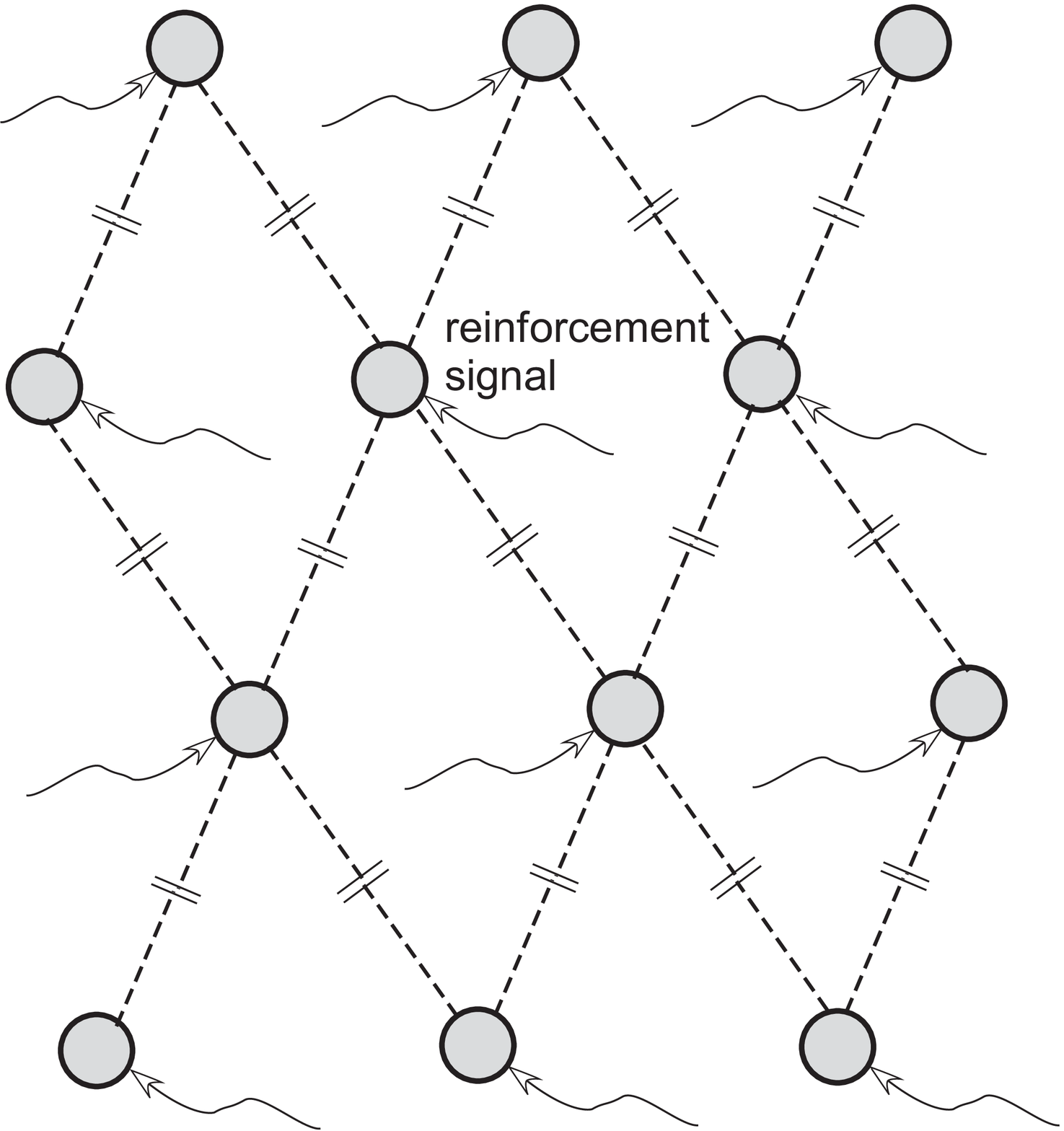}
\end{minipage}
\vspace*{8pt}
\caption{Left: Visualization of heterosynaptic plasticity experimentally found by Fitzsimonds et al.. The neurons (synapses) which are affected by heterosynaptic plasticity induced in $w_{68}$ are drawn as black circles (full lines). Right: Otmakova et al. found, that neurons in the CA1 region of the hippocampus receive a global reinforcement signal in form of dopamin.}\label{fig2}
\end{figure}

Recently, there is an increasing number of experimental results, which investigate a new form of synaptic modification, the so called heterosynaptic plasticity. In contrast to homosynaptic plasticity, where only the synapse between active pre- and postsynaptic neuron is changed, heterosynaptic plasticity concerns also further remote synapses of the pre- and postsynaptic neuron. This scenario is depicted in the left Fig. \ref{fig2}. We suppose again, that the synapse $w_{68}$ was changed either by LTP or LTD. Fitzsimonds et al. \cite{fitzsimonds_1997} found in cultured hippocampal neurons that the induction of LTD in $w_{68}$ is also accompanied by back propagation of depression in the dendrite tree of the presynaptic neuron. Further more, depression also propagates laterally in the pre- and postsynaptic neuron. Similar results hold for the propagation of LTP, see \cite{bipoo_2001} for a review. These experimental findings are depicted in the left Fig. \ref{fig2}. We emphasize all synapses, whose weights are changed $(w_{\mathrm{58}},w_{\mathrm{69}},w_{\mathrm{26}},w_{\mathrm{36}})$, and all neurons, which enclose these synapses by drawing full lines respectively black circles. A direct comparison between the left Fig. \ref{fig2}, which depicts heterosynaptic plasticity, with the right Fig. \ref{fig1}, which depicts homosynaptic plasticity, reveals the tremendous difference in the affected number of synapses and the starlike spread of plasticity in some of the synapses connected with the two neurons, which were the case for the induction of plasticity in synapse $w_{\mathrm{68}}$. We want explicitly to emphasize, that Fitzsimonds et al. found up to now no forward propagated postsynaptic plasticity. This would correspond to the synapses $w_{\mathrm{811}},w_{\mathrm{812}}$ of neuron $n_8$, which are drawn as dotted lines in the left Fig.\ref{fig2}. A biological explanation for the cellular mechanisms of these findings is currently under investigation. Fitzsimonds et al. \cite{fitzsimonds_1997} suggest the existence of retrograde signaling from the post- to the presynaptic neuron which could produce a secondary cytoplasmic factor for back-propagation and presynaptic lateral spread of LTD. On the postsynaptic side lateral spread of LTD could be explained similarly under the assumption that there is a blocking mechanism for the cytoplasmic factor which prevents forward propagated LTD. They are of the opinion that extracellular diffusible factors are of minor importance. 

The experiments of Fitzsimonds et al. \cite{fitzsimonds_1997} are certainly an extention of homosynaptic learning, which we denote briefly as Hebbian learning\footnote{We are aware, that Hebbian learning is usually only used in the context of LTP as explained in the text above and want to emphasize by this paralance that homosynaptic LTP and LTD are more interrelated than e.g. homo- and heterosynaptic LTP.}, but nevertheless both principles can be characterized as unsupervised learning because both learning types use exclusively local information available in the neural system. This is in contrast to the famous back-propagation learning rule \cite{werbos_1974,rumelharthinton1_1986} for artificial neural networks. The back-propagation algorithm is famous because until the 1980's there was no systematic method known to adjust the synaptic weights of an artificial multilayer (feed-forward) network to learn a mapping \footnote{The back-propagation algorithms was independently developed by Werbos \cite{werbos_1974} and Rumelhart et al. \cite{rumelharthinton1_1986} but became known to the broad research community after the article by Rumelhart et al..}. Still, the problem with the back-propagation algorithm is that it is not biological plausible because it requires a back-propagation of an error in the network. We emphasize, that the problem is not the back-propagation process itself, because, e.g., heterosynaptic plasticity could provide such a mechanism as depicted in the left Fig. \ref{fig2}, but the knowledge of the error, which can not be known explicitely to the neural network \cite{crick_1989}. For this reason learning by back-propagation is classified as supervised learning or learning by a teacher \cite{hertzkrogh_1991}. However, there is a modified form of supervised learning namely reinforcement learning that is biologically plausible. Reinforcement learning reduces the information provided by a teacher to a binary reinforcement signal $r$ that reflects the quality of the network's performance. Interestingly, experimental observations from the hippocampus CA1 region have shown that there is a global signal in form of dopamine  which is feedback to the neurons and causes thereby a modulation of LTD \cite{otmakhova_1998}. Schematically, this is depicted in the right Fig. \ref{fig2}. In this figure each neuron is connected with an additional edge which represents the feedback of dopamin in form of a reinforcement signal $r$. 

Based on the experimental findings by Frey et al. \cite{freymorris_1997} and Otmakova et al. \cite{otmakhova_1998}, Bak and Chialvo \cite{cb1999,bc2001} as well as Klemm et al. \cite{kbs2000} suggested biologically inspired learning rules for neural networks that combine unsupervised Hebbian (homosynaptic) with reinforcement learning. We call this kind of combination of Hebbian and reinforcement learning Hebb-like learning to indicate that the learning rule is different from Hebb, but contains nevertheless characteristics which are biological plausible. This includes the extention from purely unsupervised to a combination of unsupervised and reinforcement learning. The question which arises now is: How can one construct a Hebb-like learning rule which mimics additionally the learning behavior of heterosynaptic plasticity found by Fitzsimonds et al. \cite{fitzsimonds_1997}. This question will be addressed in the next section.

\section{The Definition of the stochastic Hebb-like learning rule}\label{def_lr}

\begin{figure*}[t!]
\centerline{
\includegraphics[width=60mm]{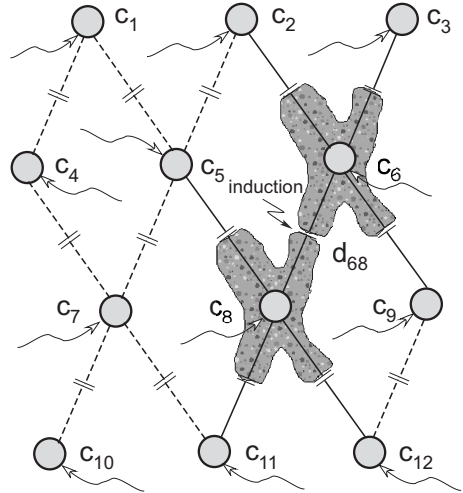}}
\vspace*{8pt}
\caption{Visualization of the symmetry of our learning rule. The cell character of a neuron is emphasized by a schematical contour around the whole neuron. Each neuron in this network has a new degree of freedom $c_i$ we call the neuron counter. Neuron counters can communicate which each other across synapses. The dynamics of the neuron counters is determined by the global reinforcement signal.}\label{fig3}
\end{figure*}

The working mechanism of the learning rule we suggest is based on the explanation of Fitzsimonds et al. \cite{fitzsimonds_1997} for heterosynaptic plasticity given above. To understand what kind of mathematical formulation is capable to describe 'a secondary cytoplasmic factor' in a qualitative way we start our explanation with emphasizing that a neuron is from a biological point of view first of all a cell. The subdivision of a neuron in synapses, soma (cell body) and axon is a model and reflects already the direction of the information flow within the neuron namely from the synapses (input) to the soma (information processing) to the axon (output). Here, we do not question this model view with respect to the direction of signal processing, but to learning. We see no biological reason why the model of a neuron for signal processing should be the same as the model of a neuron for learning. In Fig. \ref{fig3} we emphasize the cell character of a neuron by underlying the contour of the whole neuron in gray. Now, our reason for drawing the synapses in an unusual way becomes clear, because it emphasizes automatically the cell character of a neuron.

Suppose now, we assign to each neuron in the network one additional parameter $c_i$ as shown in Fig. \ref{fig3}. We call these parameters $c_i$ neuron counters. The neuron counters shall modulate the synaptic modification in a certain way defined in detail below. According to our cell view of the neuron, we assume further that the neuron counters of adjacent neurons, which are connected by synapses, can communicate with each other in an additive way. E.g., in Fig. \ref{fig3} the neuron counters $c_6$ and $c_8$ form a new value $d_{68}=c_6+c_8$ in synapse $w_{68}$, which we call the approximated synapse counter. By this mechanism we obtain a star-like influence of, e.g., the neuron counters $c_6$ and $c_8$ on all synapses connected with neuron $6$ or $8$, because either $d_{6k}=c_6+c_k$ or $d_{k8}=c_k+c_8$ holds and regulates the synaptic update of the corresponding synaptic weight of the synapses $w_{6k}$ and $w_{k8}$ respectively. This situation corresponds in a qualitative way to the learning behavior of heterosynaptic plasticity, however, with the difference, that we have a fully symmetrical learning rule. An interpretation of the communication between adjacent neuron counters can be given, if one views the neuron counters as cytoplasmic factors, which are allowed to freely move within the cytoplasm of the corresponding neuron (cell). Because, we introduced no blocking mechanism for the forward propagation of the postsynaptic neuron counter we result in a fully symmetric communication between adjacent neuron counters.

In the next section, we define the qualitative principle for heterosynaptic learning presented above mathematically. Unfortunately, there are no experimental data available that would allow to specify the influence of $d_{ij}$ on the corresponding synapse $w_{ij}$ quantitatively. For this reason, we use an ansatz to close this gap and make it plausible \cite{vphd_2003}.

\subsection{Mathematical Definition of the learning rule}\label{math_def}

If one assumes, that the neuron counters shall modulate learning, then it is plausible to determine the values of $c_i$ as a function of a reinforcement signal $r$ reflecting the performance of the network qualitatively. In the most simple case, the dynamics of the neuron counters depends linearly on the reinforcement signal.
\begin{eqnarray}
\label{synmem2}
c_{i}{\rightarrow}c'_{i}=\left\{ \begin{array}{c@{\quad}l}
{\Theta},  &c_{i}-r>{\Theta}   \\c_{i}-r,  & {\Theta}\ge c_{i}-r\ge 0\\ 0,  & 0>c_{i}-r.
\end{array}\right.
\end{eqnarray}
Here, ${\Theta}\in \mathds{N}$ is a threshold that restricts the possible values of the neuron counters $c_i$ to $\Theta+1$ possible values $\{0,\dots,\Theta\}$. The value of $c_i$ reflects the network's performance, but it has only relative and no absolute meaning with respect to the mean network error. This can be seen by the following example\footnote{We omit the index for simplicity.}. Suppose $c=0$, then it is clear that at least the last output of the network was right, $r=1$. However, we know nothing about the outputs which occurred before the last one. E.g., the following two sequences of reinforcement signals can lead to the same value of the neuron counter $c=0$: $r_1=\{1,-1,1,-1,1,-1,1\}$ and $r_2=\{1,1,1,1,1,1,1\}$ if the start value is $c=1$ for $r_1$ and $c=7$ for $r_2$. Obviously, the estimated mean error is different in both cases, if averaged over the last seven time steps. The crucial point is, that the start value of the neuron counter is not available for the neuron and, hence, the neuron can not directly calculate the mean error of the network. However, we can introduce a simple assumption, which allows an estimate of the mean network error. We claim that, if $c^1<c^2$ for one neuron in a network, but trained by two different learning rules\footnote{The superscript indicates here not the neuron in the network, but the learning rule used.} then the mean error of network one is lower then of network two. This may not hold for all cases, but it is certainly true in average. By this we couple the value of the neuron counter to the mean error of the network. Due to the fact, that this holds only statistically, we will introduce a stochastic rather than a deterministic update rule for the synapses that depends on the neuron counters. In the previous section we said, that adjacent neuron counter can communicate, if both neurons are connected by a synapse. This gives a new variable
\begin{equation}
d_{ij}=c_i+c_j
\end{equation}
we call the approximated synapse counter. We will use the approximated synapse counter as the driving parameter of our stochastic update rule, because its value reflects the performance of the synapse in the network which shall be updated, because the synapses are the adaptive part of a neural network. Hence, evaluating the value of an approximated synapse counter of a synapse will give us indirectly a decision for the update of this synapse. It is clear that, roughly speaking, the higher the approximated synapse counter of a synapse is the higher should be the probability the synapse is updated. This intuitively plausible assumption will now be quantified.

Similar to \cite{bc2001,cb1999,kbs2000} only active synapses $w_{ij}$ which were involved in the last signal processing step can be updated, if the output of the network was wrong. This is plausible, because it prevents that already learned mappings in the neural network are destroyed possibly. If $r=-1$ the probability, that synapse $w_{ij}$ is updated is given by
\begin{equation}
P_{\mathrm \Delta w}(w_{ij})=P(p^\mathrm{c}<p_\mathrm{d_{ij}}^\mathrm{r}). \label{supdate}
\end{equation}
This probability has to be calculated for each synapse $w_{ij}$ in the network. We want to emphasize, that this needs only local information besides the reinforcement signal. Hence, it is a biologically possible mechanism. If the synapse is actually chosen for update, the synaptic weight will be modified by
\begin{equation}
\label{synw1}
w_{ij}{\rightarrow}w'_{ij}=w_{ij}-{\delta}.
\end{equation}
Here, $\delta$ is a positive constant which determines the amount of the synaptic depression. To evaluate the stochastic update condition Eq. \ref{supdate} the two auxiliary variables $p^\mathrm{c}$ and $p_\mathrm{d_{ij}}^\mathrm{r}$ have to be identified. This is done in the following way:
\begin{enumerate}
\item Calculate the approximated synapse counter 
\begin{equation}
d_{ij}=c_i+c_j \label{approx_sync}.
\end{equation}
\item Map the value of the approximated synapse counter $d_{ij}$ to $p_{d_{ij}}^\mathrm{r}$ by 
\begin{eqnarray}
k&=& 2\Theta+3-d_{ij} \\
P_{k}^r  &\propto&  k^{-\tau},\; \tau  \in  \mathds{R}^+,\; k  \in  \{1,\dots  ,2\Theta+3\}.
\label{p_rank}
\end{eqnarray}
We call $P_{k}^r$  rank ordering probability distribution 
\item The random variable $p^\mathrm{c}$ is drawn from the continuous coin distribution
\begin{equation}
P^{c}(x)  \propto  x^{-\alpha},\; \alpha  \in  \mathds{R}^+,\; x \in (0,1].
\label{p_coin}\\
\end{equation}
\end{enumerate}
We had three reasons to choose a power law in Eq. \ref{p_coin} for the coin distribution instead of an equal distribution, which would be the simplest choice. First, we see no evidence that a random number generator occurring in a neural system should favor a equal distribution. Second, it is highly probable that two different random number generators of the same biological system are not identical. Instead, they could have different parameters, in our case they could have different exponents. In this paper we will content ourself investigating the case of identical random number generators, but our framework can be directly applied to the described scenario. Third, by choosing $\alpha=0$, the coin distribution in Eq. \ref{p_coin} becomes the equal distribution. This allows us to investigate the influence of the distance of the coin distribution to an equal distribution on the learning behavior of a neural network by studying different parameters of $\alpha$. We want to remark, that in this case the update probability Eq. \ref{supdate} simplifies to
\begin{equation}
P_{\mathrm \Delta w}(w_{ij})=p_\mathrm{d_{ij}}^\mathrm{r}.
\end{equation}

\begin{figure*}[t!]
\centerline{
\includegraphics[width=95mm]{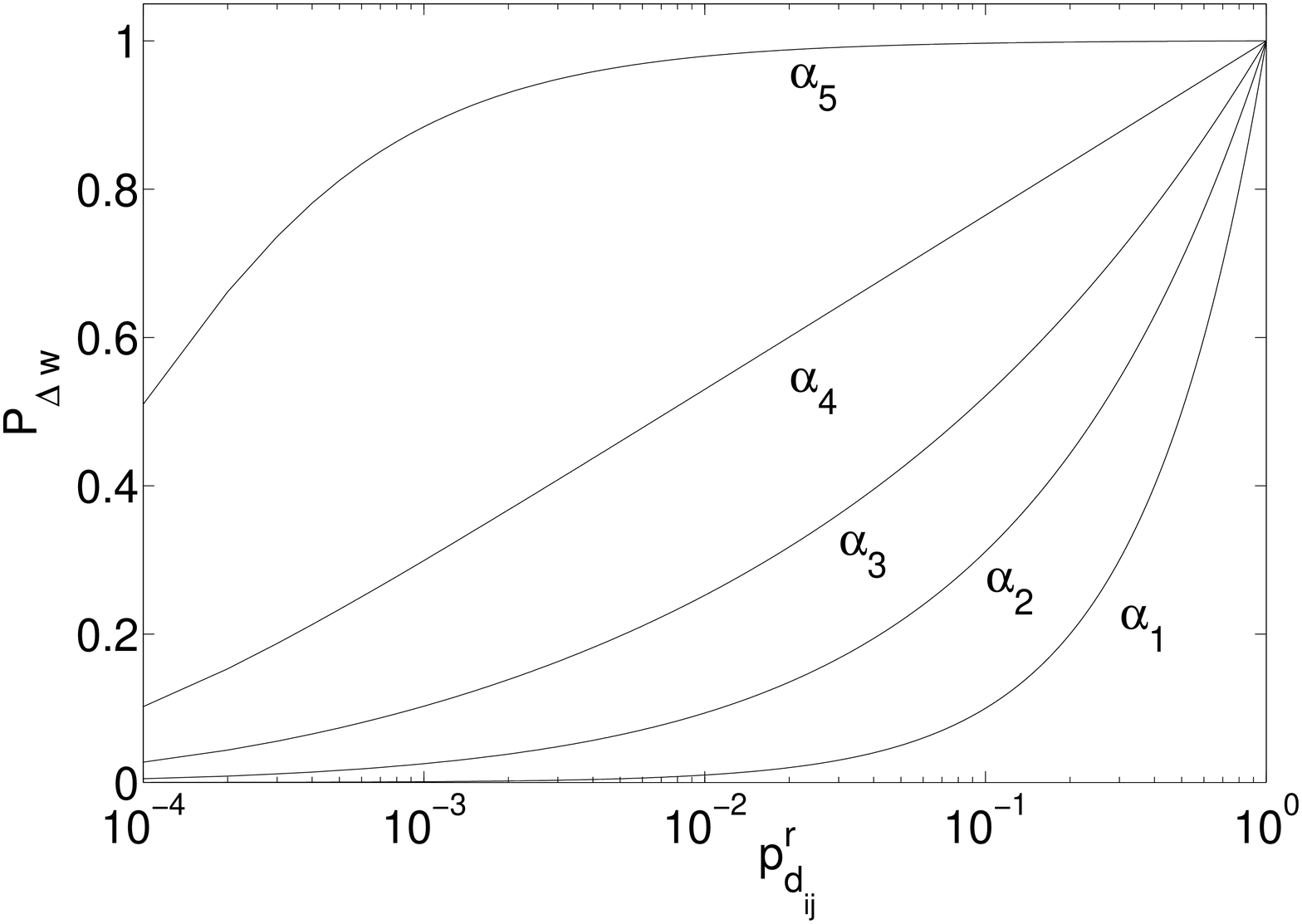}}
\vspace*{8pt}
\caption{Update probability $P_{\mathrm \Delta w}$ as function of $p_\mathrm{d_{ij}}^\mathrm{r}$. The different curves correspond to different values of the exponent $\alpha$ of the coin distribution. The values are: $\alpha_1=0.0$, $\alpha_2=0.5$, $\alpha_3=0.75$, $\alpha_4=1.0$ and $\alpha_5=1.75$.}\label{fig3_new}
\end{figure*}
\begin{figure*}[t!]
\centering
\begin{minipage}[c]{0.49\textwidth}
\includegraphics[width=63mm]{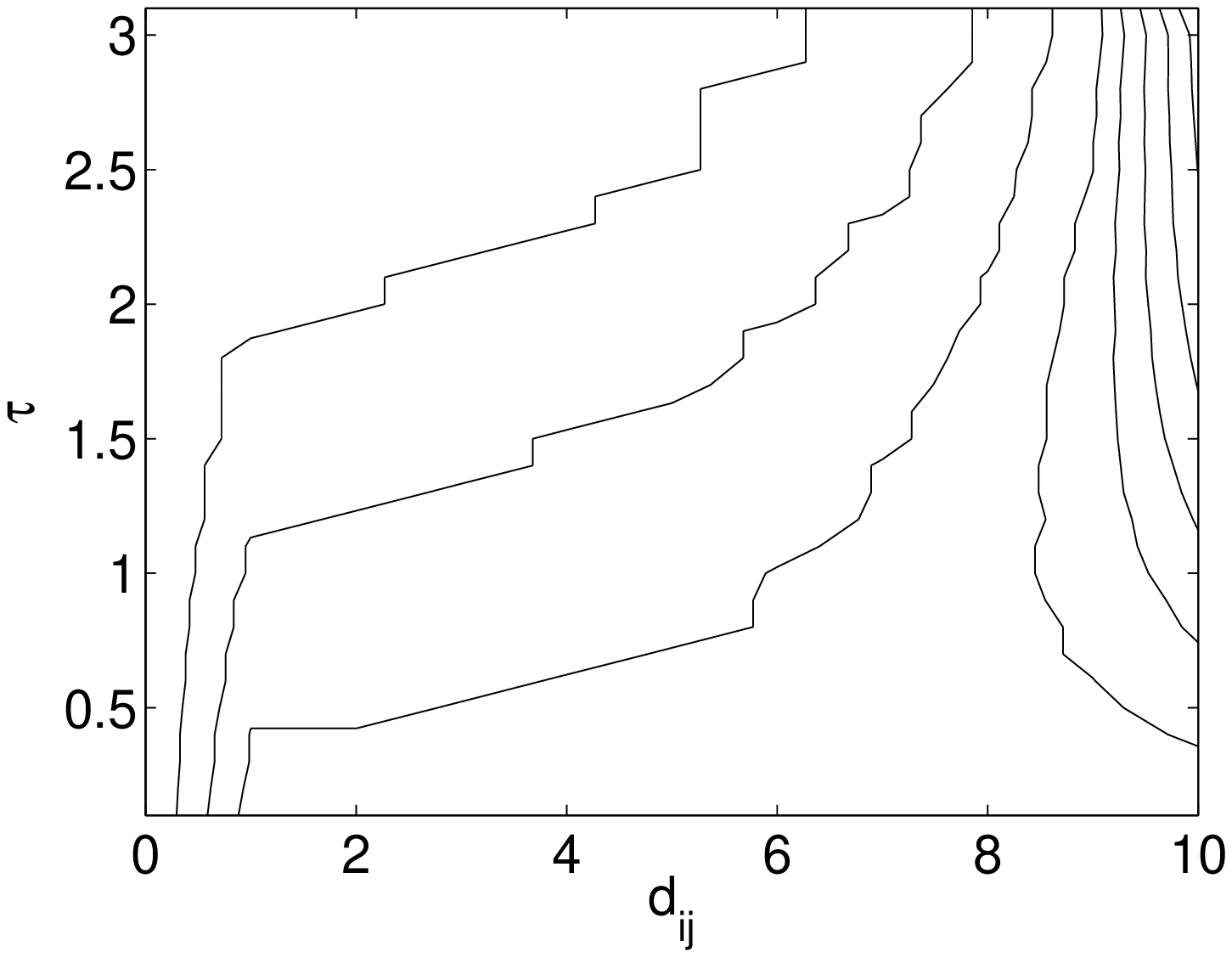}
\end{minipage}
\centering
\begin{minipage}[c]{0.49\textwidth}
\includegraphics[width=63mm]{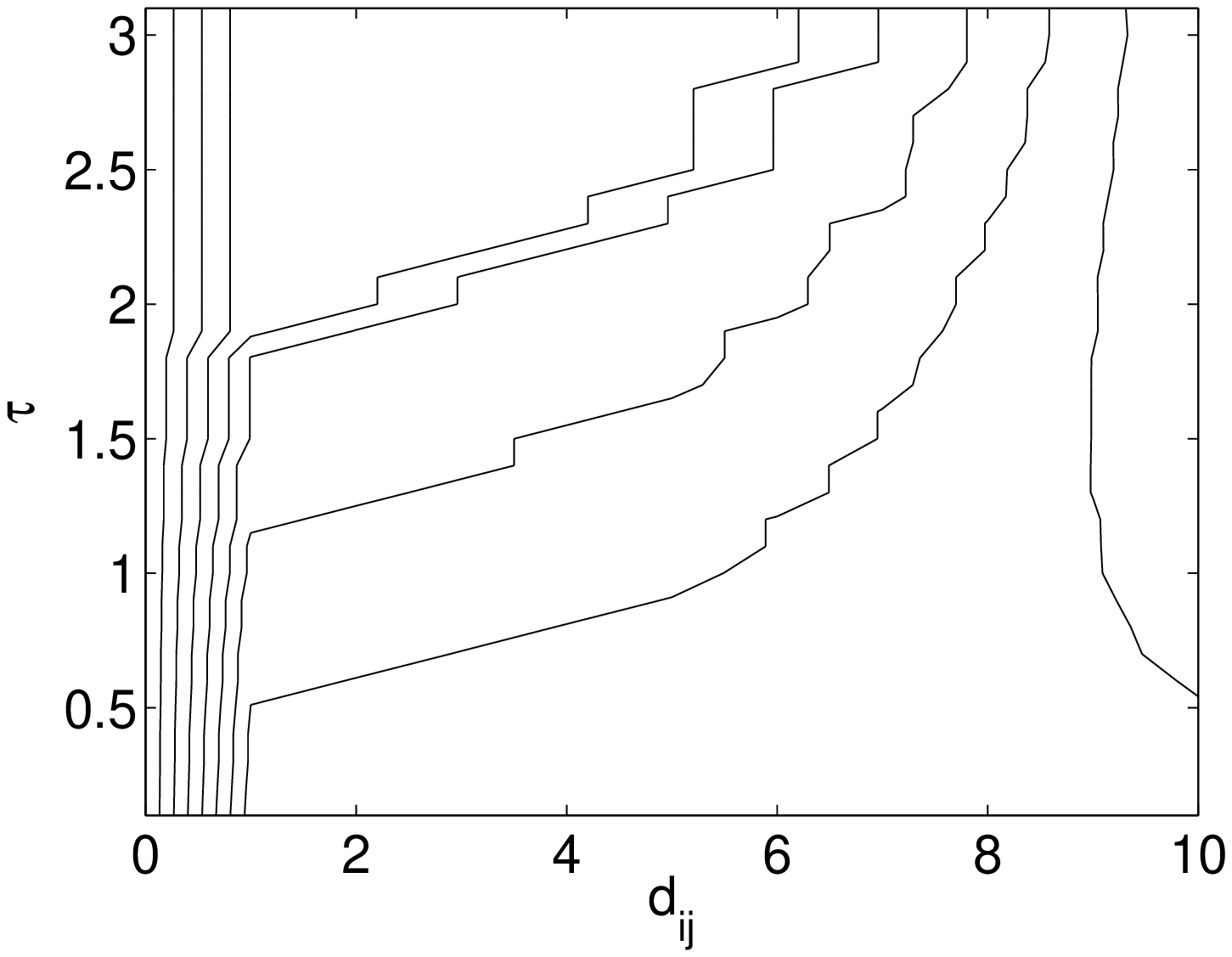}
\end{minipage}
\vspace*{8pt}
\caption{Update probability $P_{\mathrm \Delta w}$ as function of $\tau$ and $d_{ij}$ obtained for $\Theta=4$. The contour plot of $P_{\mathrm \Delta w}$ is shown. The update probability is close to one in the upper right corner and close to zero in the uppler left corner. The exponent $\alpha$ of the coin distribution was $\alpha=0.7$ in the left and $\alpha=1.5$ in the right figure. }\label{fig_update2}
\end{figure*}

Before we present our results in the next section, we want to visualize the stochastic update probability $P_{\mathrm \Delta w}$. Figure \ref{fig3_new} shows the update probability $P_{\mathrm \Delta w}$ as function of $p_\mathrm{d_{ij}}^\mathrm{r}$. The different curves correspond to different values of the exponent $\alpha$ of the coin distribution. One can see, that the update probability follows the values of $p_\mathrm{d_{ij}}^\mathrm{r}$. This holds for each curve in Fig. \ref{fig3_new}. That means, the higher the values of $p_\mathrm{d_{ij}}^\mathrm{r}$ are the higher is the update probability. This is the behavior one would intuitively expect, because high values of $p_\mathrm{d_{ij}}^\mathrm{r}$ correspond to high values of the approximated synapse counters $d_{ij}$ indicating high values of the neuron counters, which correspond to a bad network performance. Moreover, one can see in Fig. \ref{fig3_new} that the larger $\alpha$ the higher is the update probability for fixed $p_\mathrm{d_{ij}}^\mathrm{r}$. In the limit $\alpha \rightarrow \infty$ the update probability equals one for all values of $p_\mathrm{d_{ij}}^\mathrm{r}$. Hence, higher values of the exponent $\alpha$ of the coin distribution result in a higher update probability. That means, by $\alpha$ one can control the sensitivity by which the update probability depends on $p_\mathrm{d_{ij}}^\mathrm{r}$. Another parameter our stochastic update rule depends on is the exponent of the rank ordering distribution $\tau$. We display in Fig. \ref{fig_update2} $P_{\mathrm \Delta w}$ as function of $\tau$ and $p_\mathrm{d_{ij}}^\mathrm{r}$ to visualize its influence on the update probability. The values of the update probability are color-coded and blue corresponds to $0$ and red to $1$. For the left Fig. \ref{fig_update2} we used $\alpha=0.7$ and for the right $\alpha=1.5$ as exponent for the coin distribution. If $p_\mathrm{d_{ij}}^\mathrm{r}=0$ no update takes place. For increasing values of the approximated synapse counter and fixed values of $\tau$ one obtains increasing values for the update probability. Moreover, higher values of $\alpha$ lead to higher values of $P_{\mathrm \Delta w}$. This can be seen by comparing the left and right Fig. \ref{fig_update2}. Increasing values of $\tau$ result in decreasing values of $P_{\mathrm \Delta w}$ for fixed $d_{ij}$. 
To summarize, the stochastic update condition we introduced for a synaptic update depends on six parameters
\begin{equation}
P_{\mathrm \Delta w}(w_{ij})=P_{\mathrm \Delta w}(r,c_i,c_j,\Theta,\alpha,\tau).
\end{equation}
From the visualizations we gave in Fig. \ref{fig3_new} and \ref{fig_update2} we saw that increasing values of $d_{ij}$ and $\alpha$ as well as decreasing values of $\tau$ lead to an increase in the update probability.

%%%%%%%%%%%%%%%%%%%%%%%%%%%%%%%%%%%%%%%%%%%%%%%%%%%%%%%%%%%%%%%%%%

%%%%%%%%%%%%%%%%%%%%%%%%%%%%%%%%%%%%%%%%%%%%%%%%%%%%%%%%%%%%%%%%%%%%%%%

\section{Numerical Simulations}\label{results}

For the following simulations we use a three-layer feed-forward network. The neural network consist of $I$ input-, $H$ hidden- and $O$ output neurons. The neurons of adjacent layers are all to all connected with synapses $w_{ij}\in\mathds{R}$. As neuron model we us binary neurons $x_i\in\{0,1\}$ for $i\in\{1,\ldots,I+H+O\}$. The network dynamics is regulated by a winner-take-all mechanism whereas the inner fields of the neurons are calculated by 
\begin{equation}
h_j=\sum^\mathrm{all}_i w_{ji}x_i.\label{innerfield}
\end{equation}
Here, {\em{all}} means all neurons of the preceding layer. As active neuron in each layer we choose the neuron with the highest activity
\begin{equation}
i_\mathrm{max}=\mathop{\mathrm{argmax}}_{i}(h_i)\label{argmax}
\end{equation}
which is set to $x_{i_{\mathrm{max}}}=1$. All other neurons are set to zero. By this we enforce a sparse coding. Bak and Chialvo \cite{cb1999} have called this {\em{extremal dynamics}}.

The training of the neural network works as follows: We choose randomly one of the possible input patterns and initialize the neurons in the input layer. Then we calculate according to the network dynamics Eq. \ref{innerfield}-\ref{argmax} the activity of the neurons in the subsequent layers. If the output of the network is correct we set $r=1$ otherwise the reinforcement signal is set to $r=-1$. According to Eq. \ref{synmem2} we calculate the new values of the neuron counters for the neurons which were active during the signal processing of the input pattern. If $r=-1$ we apply our stochastic learning rule otherwise we proceed with the next input pattern until the network converged.

The mapping which shall be learned by the network is the exclusive-or (XOR) function and higher dimensional extensions thereof called the parity problem. One can describe the mappings from the parity problem class as indicator functions for an odd or even number of $1$'s in the binary input vector $(x_1^I,\dots,x_k^I)$ of the network. If the number of $1$'s in the input vector is odd the output of the network shall be $(x_1^O=1,x_2^O=0)$ if it is even $(x_1^O=0,x_2^O=1)$. In this sense, the exclusive-or (XOR) function is the two dimensional $k=2$ representative of this class. To avoid the case of a zero input vector, which would result in zero activity of subsequent layers, we introduce a bias neuron $x_{k+1}^I=1$. Here, the index $k$ is given by the exponent of the maximal number of patterns $p=2^k$ which can be realized by a random binary vector of length $k$. For the following simulations the initial weights of the network were chosen randomly from $[0,1]$ and the neuron counters were all set to zero. The learning rate $\delta$ was randomly chosen from $[0,\delta_0]$, with $\delta_0=0.1$, each time when a synaptic modification was induced.

%%%%%%%%%%%%%%%%%%%%%%%%%%%%%%%%%%%%%%%%%%%%%%%%%%%%%%%%%%%%%%%%%%%%%%%

%%%%%%%%%%%%%%%%%%%%%%%%%%%%%%%%%%%%%%%%%%%%%%%%%%%%%%%%%%%%%%%%%%%%%%%

\subsection{XOR Function}\label{wta}

\begin{figure}[t!]
\begin{center}
\begin{tabular}{c}
\includegraphics[width=60mm]{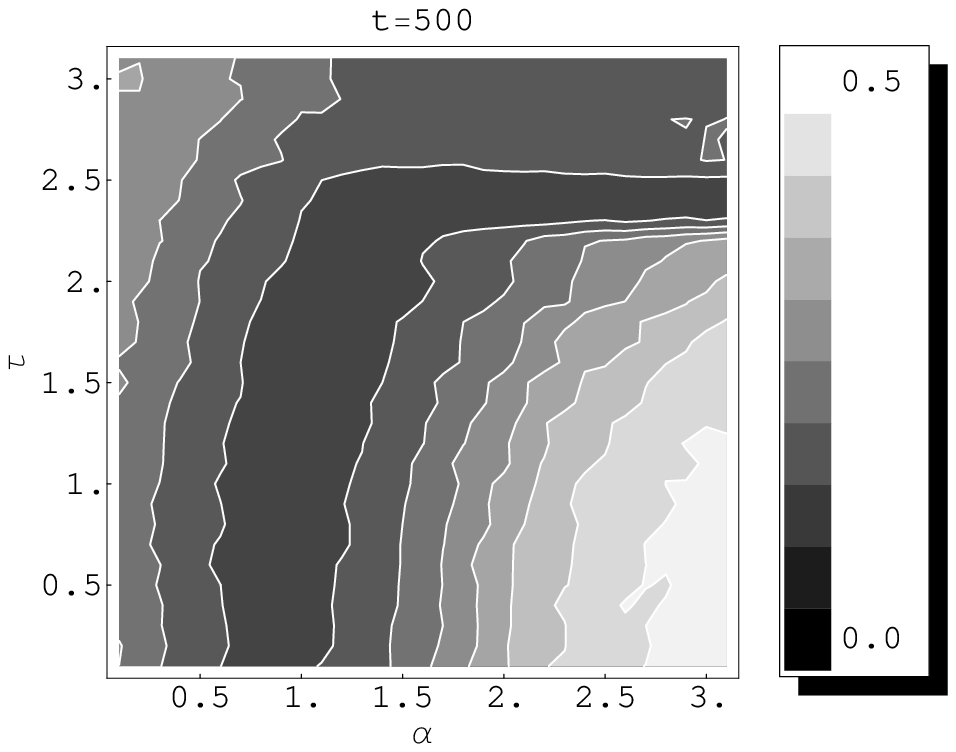}
\includegraphics[width=60mm]{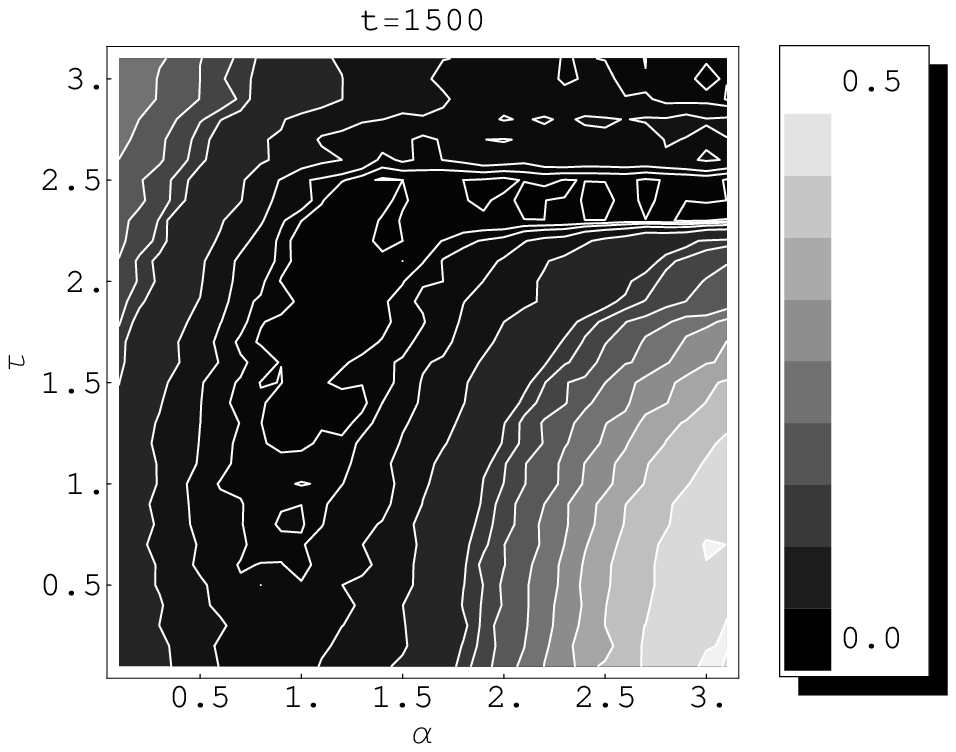}
\end{tabular}
\end{center}
\vspace*{8pt}
\caption{Mean ensemble error $E$ in dependence on $\tau$ and $\alpha$ at the time steps $t=500$ (left) and  $t=1500$ (right) during the learning process for $\Theta=3$. The network dynamics was a winner-take-all mechanism and the ensemble size $10000$.}\label{fig_res1}
\end{figure}

We start our investigations by studying the influence of the memory length of the neuron counters $\Theta$ and the exponents $\alpha$ and $\tau$ on the mean ensemble error $E$ of the network's performance during learning the XOR function. The contour plot in Fig. \ref{fig_res1} shows the simulation results for $\Theta=3$ and three neurons in the hidden layer. The mean ensemble error $E$ was obtained by averaging over independent runs of an ensemble of size $10000$ and is displayed at the time steps $t=500$ (left figure) and $t=1500$ (right figure) during the learning process. To find the optimal parameter configuration
\begin{eqnarray}
(\Theta^*,\alpha^*,\tau^*)=\mathop{\mathrm{argmin}}_{\Theta,\alpha,\tau} E(\Theta,\alpha,\tau;t)
\end{eqnarray}
 which minimizes the mean ensemble error $E(\Theta,\alpha,\tau;t)$ we keep $\Theta$ fixed and vary $\alpha$ and $\tau$ in the interval $[0.0,3.0]$ in $10^{-1}$ steps. 
\begin{table}[b!]
\tbl{Minimal mean ensemble error $E_\mathrm{min}(\Theta^*,\alpha^*,\tau^*;t)$ obtained by the optimal parameters $\Theta^*$,$\tau^*$ and $\alpha^*$ for the time steps $t=500,1000$ and $1500$ (left, middle and right column). The ensemble size for each simulation was $10000$.}
{\begin{tabular}{c|ccc|ccc|ccc} \hline
$\Theta^*$&&$\tau^*$&&&$\alpha^*$&&&$E_\mathrm{min}$&\\ \hline
1 &2.1&2.0&2.2&0.6&0.6&0.8&0.099&0.021&0.004  \\
2 &1.8&2.1&2.6&0.9&1.1&1.2&0.113&0.122&0.007  \\
3 &2.4&2.3&2.3&1.7&1.7&1.7&0.122&0.032&0.009 \\
4 &2.2&2.2&2.2&2.8&3.0&3.0&0.112&0.022&0.004  \\
5 &1.9&1.9&1.9&2.4&3.0&3.0&0.087&0.018&0.003 \\\hline \hline
t & 500 & 1000 & 1500 & 500 & 1000 & 1500 & 500 & 1000 & 1500\\ \hline
\end{tabular}
\label{tab1}}
\end{table}
\begin{figure*}[t!]
\begin{center}
\begin{minipage}[c]{0.50\textwidth}
\includegraphics[width=90mm]{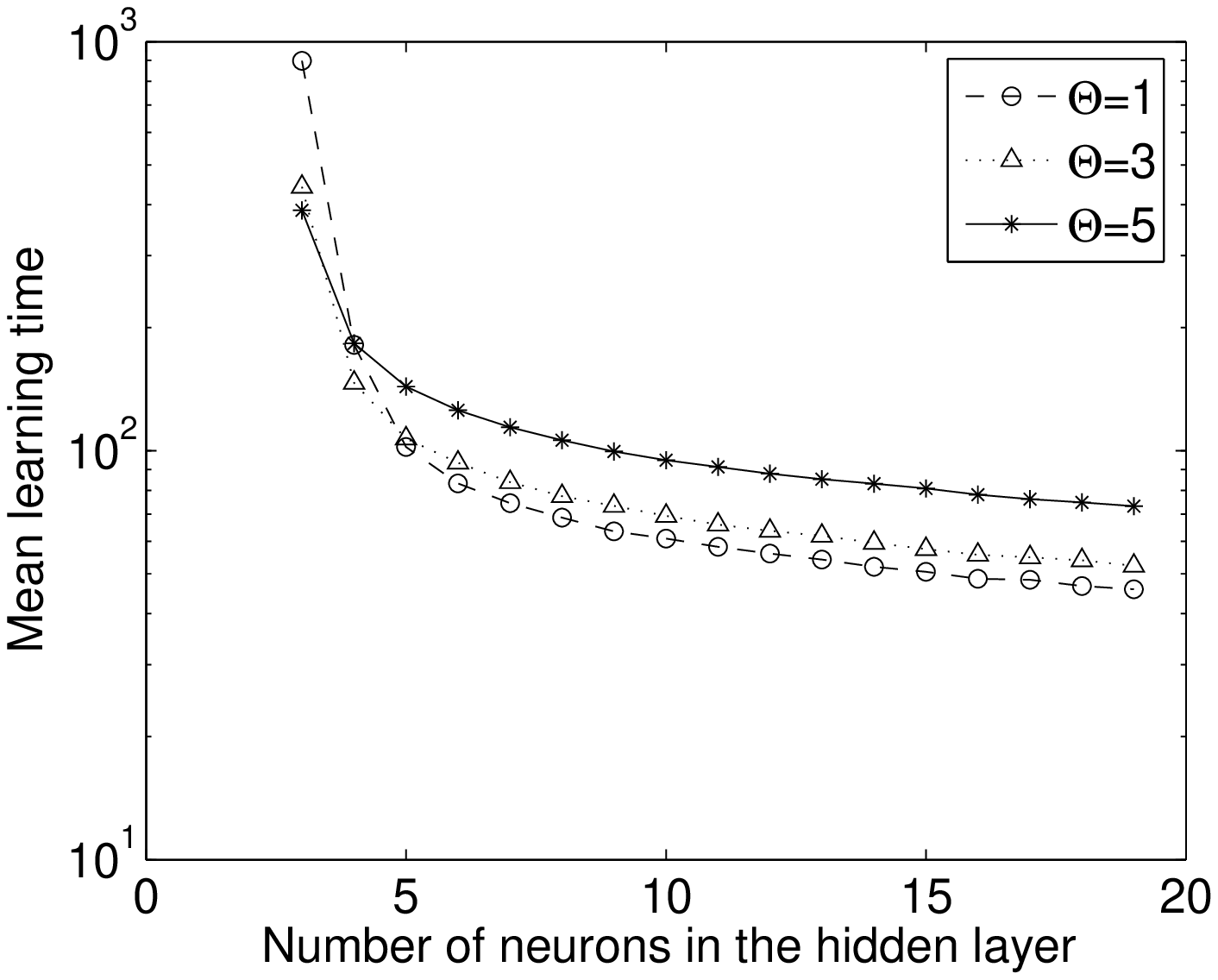}
\end{minipage}
\begin{minipage}[c]{0.50\textwidth}
\includegraphics[width=90mm]{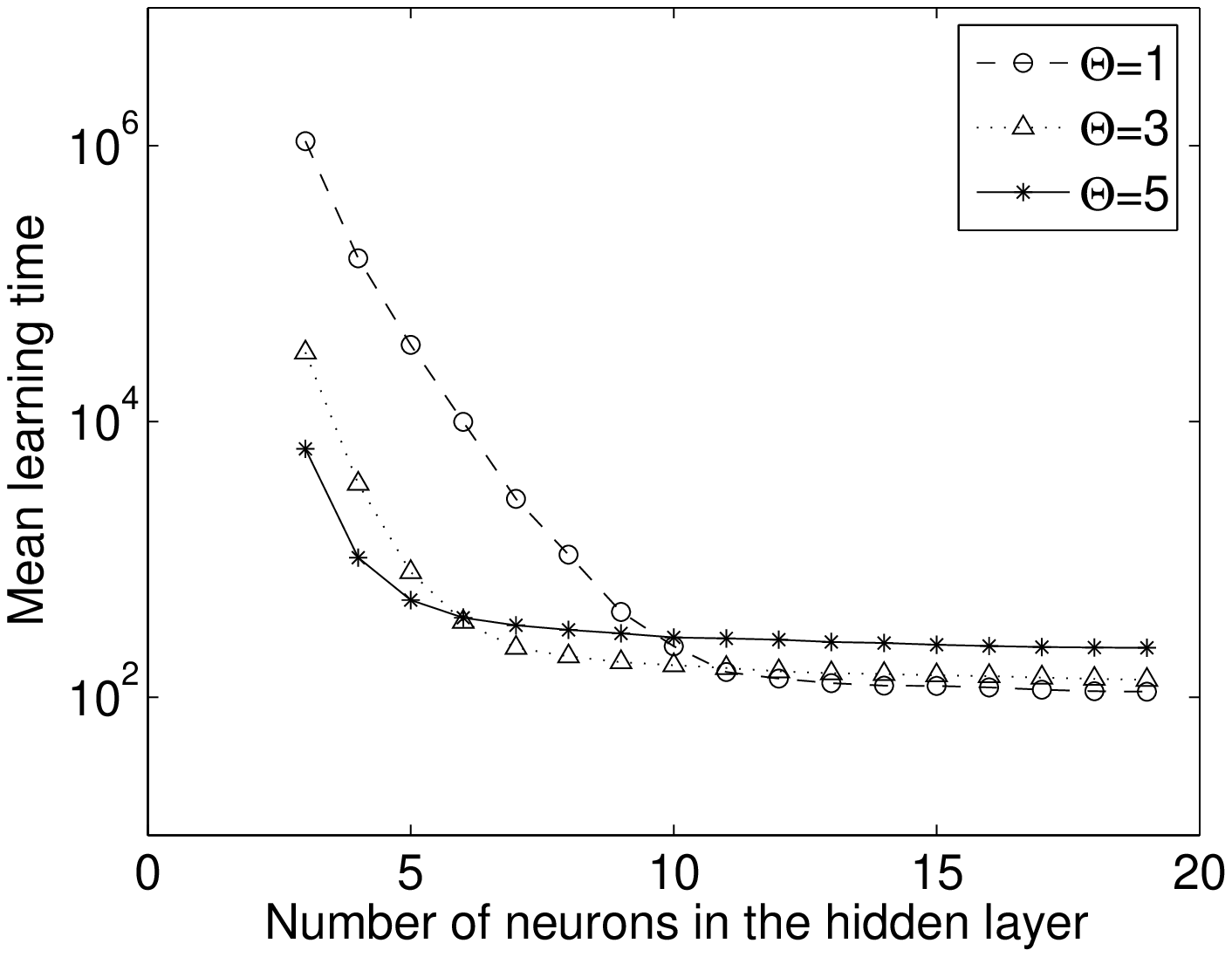}
\end{minipage}
\end{center}
\caption{Mean learning time in dependence on the number of neurons in the hidden layer. $\tau$ and $\alpha$ are given in table \ref{tab1} at the time step $t=1500$ for the corresponding values of $\Theta$. The network dynamics was a winner-take-all mechanism (upper figure) and a noisy winner-take-all mechanism (lower figure). The ensemble size was for all simulations $10000$.}\label{fig_res2}
\end{figure*}

From Fig. \ref{fig_res1} one can see that learning takes place in the whole parameter space $(\alpha,\tau)$. Of course there are regions in which learning is much faster than in others due to the fact that the resulting update probability of our learning rule, controlled by $(\alpha,\tau)$, is more suitable for the learning task. To investigate the $\Theta$ dependence of our learning rule we repeated these simulations for several $\Theta$ values. The results for the optimal parameter configurations $(\Theta^*,\alpha^*,\tau^*)$ from these simulations can be found in table \ref{tab1}. From these results one can conclude that there is no single parameter configuration in this $3$ dimensional parameter space, which minimizes $E$. But there exist multiple parameter configurations resulting in almost the same performance with respect to the absolute convergence of the network. Interestingly, from table \ref{tab1} one can see, that with increasing values of $\Theta$, $\alpha$ also increases but $\tau$ is almost constant. Based on our explanation in section \ref{math_def} about the dependence of $P_{\mathrm \Delta w}$ on $\alpha$ and $\tau$ we can conclude, that higher values of $\Theta$ require a higher update probability. This makes sense, because the complexity of the mapping to be learned by the network was not changed. Only the memory length of the neuron counters was enlarged. Apparently, this was not necessary and, hence, would result in worse results, because averaging over a longer time interval $\Theta$ is more time consuming. This effect is compensated by the higher $\alpha$ value resulting in more frequent updates. For our subsequent investigations we use the optimal parameter values obtained at the learning time step $t=1500$ from table \ref{tab1}.

Based on these results we study systematically the dependence of the mean learning time from the network topology and the network dynamics. In the left Fig. \ref{fig_res2} we show the mean learning time as function of the number of neurons $H$ in the hidden layer. The curves are indexed by different values of the neuron counter $\Theta$. In the lower figure we demonstrate the robustness of these results in the presence of noise $\eta$ by using a noisy winner-take-all mechanism as network dynamics which adds to the inner fields Eq. \ref{innerfield} of the neurons noise $\eta$ before the neuron with the highest inner field is selected. The noise was uniformly drawn from $[0,\eta_0]$ with $\eta_0=\frac{\delta_0}{2}$. From both figures one can see that the mean learning time decreases with an increasing number of neurons in the hidden layer as expected whereas the increase from $3$ to $4$ neurons has the biggest effect. This is due to the fact that the destructive path inference, which means that already correctly learned paths in the network are destroyed by a new synaptic modification, is strongly reduced by increasing the number of possible paths as a result of additional neurons in the hidden layer. Increasing the number of neurons beyond $19$ has only marginal influence because an additional increase of redundant paths has no affect. Even in the presence of noise our learning rule is capable of learning the XOR function. One can nicely see how an increasing number of neurons in the hidden layer can efficiently reduce the amount of noise in the system.

\subsection{k-dimensional parity functions}

In this subsection we study the influence of the number of patterns to be learned on the mean learning time. We use $p=2^k$ input patterns, for $k\in{1,\dots,6}$, and correspondingly $I=k+1$ neurons in the input layer\footnote{This includes one neuron as bias.} and $H=1500$ neurons in the hidden layer. 
\begin{figure}[t!]
\centerline{
\includegraphics[width=85mm]{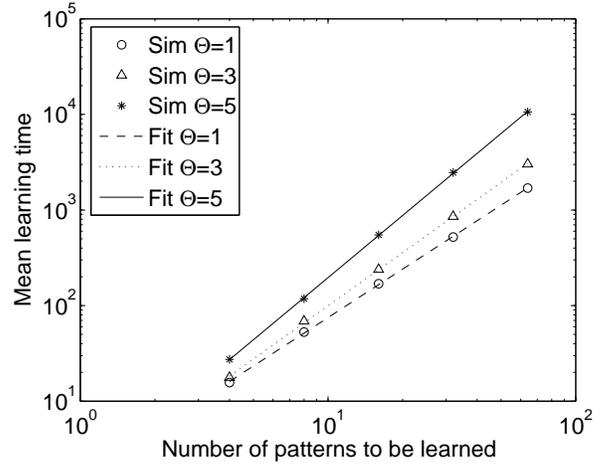}}
\vspace*{8pt}
\caption{Mean learning time in dependence on the number of patterns to be learned. The network consisted of $H=1500$ neurons in the hidden layer. The mean learning time was averaged over an ensemble of size $1000$. The symbols correspond to results obtained from simulations whereas the lines are the results from a least mean square fit. The exponents for the power laws are $\beta=\{1.68,1.84,2.15\}$ in acceding order of $\Theta$.}\label{fig_res3}
\end{figure}
The network dynamics was again regulated by a winner-take-all mechanism. Our results shown in Fig. \ref{fig_res3} for the mean learning times are comparable to the results obtained by Bak and Chialvo \cite{bc2001} with the difference that they even used $3000$ neurons in the hidden layer. Moreover, the mean learning time scales\footnote{See caption to figure \ref{fig3} for numerical values for $\beta$ for the three different curves.} with problem size $p$ according to a power law $\sim p^{\beta}$ with exponent $\beta\sim 1.8$. This demonstrates not only, that our stochastic learning rule is able to learn the problem but also, that learning is efficient, because otherwise the mean learning times would follow an exponential function.

\subsection{Influence of exponential distributions on the learning behavior}

\begin{figure*}[h!]
\centering
\includegraphics[width=80mm]{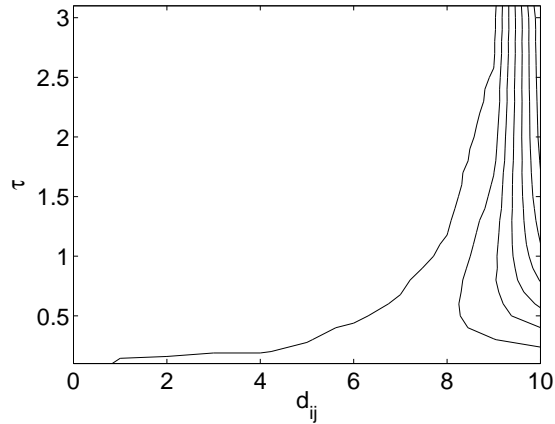}
\vspace*{8pt}
\caption{Contour plot of the update probability $P_{\mathrm \Delta w}$ as function of $\tau$ and $p_\mathrm{d_{ij}}^\mathrm{r}$ obtained for $\Theta=4$. The parameter of the coin distribution was $\alpha=1.0$. The coin and the rank ordering distribution was an exponential function. The update probability is almost always close to zeros but increases rapidly for $\tau \rightarrow 3 $ and $d_{ij} \rightarrow 10$.}\label{fig_exp}
\end{figure*}

Finally, we investigated the influence of the type of the probability distribution used for the coin and rank ordering distribution. Here, we use an exponential distribution for the coin and rank ordering distribution and study the learning behavior. We found significantly worse results compared to the results for the power law (not shown) presented in the last section. To understand this, we display in Fig. \ref{fig_exp} the update probability as function of $\tau$ and $d_{ij}$.
One can see, there are essentially only two states, the update probability can take, zero and one (upper right). That means, $P_{\mathrm \Delta w}$ produces a rather deterministic update behavior which is inappropriate, because the information provided by the approximated synapse counters is uncertain. Other values for $\alpha$ show qualitatively the same results. This demonstrates that the larger variability provided by a power law distribution is important for a good learning behavior.

%%%%%%%%%%%%%%%%%%%%%%%%%%%%%%%%%%%%%%%%%%%%%%%%%%%%%%%%%%%%%%%%%%%%%%%%

\section{Discussion and comparison of learning rules}\label{discussion} 

Mathematical investigations of biological as well as artificial learning rules for neural networks have been attractive to scientists since decades, because of the importance of the underlying problem and implications arising out of an understanding thereof. We want to finish this article, by discussing and comparing our novel stochastic Hebb-like learning rule with other models introduced so far, which are constrained in a way that makes them biologically plausible.

Bak and Chialvo \cite{cb1999,bc2001} introduced a learning rule which combines Anti-Hebb or long-term depression (LTD) and reinforcement learning. Klemm et al. \cite{kbs2000} extended the learning rule from Bak and Chialvo by introducing one additional degree of freedom for each synapse in the network. They called this degree of freedom synapse counter. Moreover, Bosman et al. proposed a learning rule which incorporates Hebb (LTP), Anti-Hebb (LTD) and reinforcement learning \cite{bosman_2004}. All these approaches have in common with our learning rule, that they utilize a reinforcement signal as feedback reflecting the current performance of the network. The usage of a reinforcement signal seems not only to be plausible but indispensable to learn mappings, because the neural network has to adapt to its environment by interacting with it otherwise the animal will die fast. Similar to physical energy\footnote{Perpeduum mobile.}, it is also impossible to generate information out of nothing in a meaningful way. The reinforcement signal makes a neural network and, hence, a brain, an open system according to the flow of information. This depicts intuitively the difficulty of the system under investigation, because open or dissipative systems are by far less understood than closed, e.g., Hamiltonian systems. 

In contrast, all models \cite{cb1999,bc2001,kbs2000,bosman_2004} proposed before are purely deterministic with respect to the decision if an update for a synapse shall take place or not. Additionally, all learning rules \cite{cb1999,bc2001,kbs2000,bosman_2004} can only explain homosynaptic plasticity. We think, due to the fact that the neural network is an open system it can not make deterministic decisions which are objective, because of the lack of complete information. Of course, one can always search for the best decision based on the amount of information available in the system. However, this internal (in the neural network) optimality does not guarantee external (the overall network performance) optimality. In this article, we took the point of view, that we assume we have incomplete information and, hence, we are only able to provide an update probability indicating a kind of confidence level for this update based on our incomplete information. Explicitely, this enters our model in form of the approximated synapse counters. For every network topology one can calculate the synapse counter as a function of the neuron counters introduced by Klemm et al. \cite{kbs2000}. However, this results normally in relations, which involve not only the neuron counters enclosing the synapse, but also further remote neuron counters \cite{vphd_2003}. This can be seen with the help of Fig. \ref{fig3}. For example, the neuron counter of neuron five can be written as a linear sum of the synapse counters:
\begin{eqnarray}
c_5 = d_{25} + d_{35} \\
c_5 = d_{57} + d_{58} \label{coding}
\end{eqnarray}
These equations represent a failure conservation for the incoming and outgoing connections respectively. If the neuron counter of neuron five is $c_5$ then the sum of all synapse counters leading to neuron five has to be equal to this number, because there is no other way information can involve neuron five in the signal processing. The same holds for the outgoing information, represented by Eq. \ref{coding}. In general, such linear failure conservation relations between the neuron and synapse counters lead to the linear system 
\begin{eqnarray}
c_n=\mathcal{M}c_s \label{coding2}
\end{eqnarray}
Here, $c_n$ represents the $N$-dimensional vector of neuron and $c_s$ the $S$-dimensional vector of synapse counters. The integer valued $N$ times $S$ matrix $\mathcal{M}$ depends on the network topology. The problem becomes nonlinear if one wants to obtain the synapse counters as function of the neuron counters, because the inverse of the non-quadratic matrix $\mathcal{M}$ in Eq. \ref{coding2} can only be done by calculating a pseudoinverse to obtain $c_s=\mathcal{M}^{-1}c_n$. This is the situation we are facing. Explicite calculation by using the Moore-Penrose pseudo inverse \cite{penrose1955} leads to the statement given above \cite{vphd_2003}. Hence, a biologically plausible learning rule can not use these relations, because this would violate the local information condition in neural networks. One possibility around this obstacle is to approximate the synapse counter by the sum of the neuron counters enclosing this synapse, however, with the additional assumption to view the resulting value in a probabilistic rather than deterministic way. Our simulations showed, that a merely addition (or multiplication) of the neuron counters does not lead to meaningful results at all \cite{vphd_2003}. Moreover, also the used probability distributions have significant influence on the learning dynamics as demonstrated in the results section \ref{results}. The fact, that power law distributions give significantly better results than exponential distributions for the coin and rank ordering distribution corresponds to results of recent investigations of heuristic optimization strategies. Boettcher et al. \cite{bp2001} demonstrated that the usage of power law distributions in optimization problems, e.g., finding the energy ground states for spin glasses \cite{bp2001} and graph bi-partitioning \cite{bp2000}, which are both NP-hard optimization problems, can give better results compared to simulated annealing \cite{kirk1983} or genetic algorithms \cite{holland1975}. They explained this effect by the positive influence of the inherently large fluctuations within the system, which prevents to get trapped a long time in local minima of the error function. 

From a biological point of view the most significant difference between our stochastic Hebb-like learning rule and the other learning rules \cite{cb1999,bc2001,kbs2000,bosman_2004} is certainly that our model aims to explain heterosynaptic plasticity, which has been found experimentally \cite{fitzsimonds_1997}, instead of homesynaptic plasticity, in a qualitative way. This is also the major objective of this paper. Hence, a direct comparison between our model and the other learning rules can not be given fairly without neglecting or underestimating significant components of our model. For example, we introduced one new degree of freedom for each neuron in the form of neuron counters. Bosman et al. \cite{bosman_2004} do not rely on this or similar parameters whereas Klemm et al. \cite{kbs2000} introduced one additional degree of freedom for each synapse. That means, in this context our model has $N$ parameters, the model of Bosman et al. none, and Klemm et al. $kN$ parameters. Here, let $k$ be the average number of synapses a neuron has in a network. This makes the learning rule of Bosman et al. in a mathematical sense minimal compared to ours. However, biologically it can not describe heterosynaptic plasticity and, hence, lacks this ability, which makes a comparison in the number of parameters meaningless. Interestingly, despite the fact, that heterosynaptic plasticity is more complex then homosynaptic plasticity the learning rule of Klemm et al. uses $k$ times more parameters than our model. In general, we think that due to the almost overwhelming complexity of biological phenomena mathematical modeling should stay always in tight contact with experimental findings to constrain the model by regularities found in nature. These constrains can only lead to minimal mathematical models in the context under consideration, but not beyond.

%%%%%%%%%%%%%%%%%%%%%%%%%%%%%%%%%%%%%%%%%%%%%%%%%%%%%%%%%%%%%%%%%%%%%%%%

\section{Conclusions}\label{end}

We presented a novel stochastic Hebb-like learning rule for neural networks and demonstrated its working mechanism exemplary in learning the exclusive-or (XOR) problem in a three-layer network. We investigated the convergence behavior by extensive numerical simulations depending on three different network dynamics which correspond all to biological forms of lateral inhibition. We found in all cases parameter configurations for $\Theta$, the length of the neuron memory, $\alpha$, the exponent of the coin distribution and $\tau$, the exponent of the rank ordering distribution, which constitute the Hebb-like learning rule, to obtain not only a solution to the exclusive-or (XOR) problem but comparably well results to a learning rule recently proposed by Klemm, Bornholdt and Schuster \cite{kbs2000}. This is remarkable, if one keeps in mind that our learning rule uses less parameters than the model proposed by \cite{kbs2000}. Because the number of neurons is always (much) less then the number of synapses the same holds for the respective numbers of synaptic and neuron counters which were used in the learning rules.

An interesting implication of our learning rule and its inherent stochastic character is that it offers a quantitative biologically plausible explanation of heterosynaptic plasticity which is observed experimentally. In addition to the experimentally observed back-propagation, pre- and postsynaptic lateral spread of {\em{long-term depression}} (LTD) our learning rule predicts forward propagated postsynaptic LTD for reasons of a symmetric communication between adjacent neurons. As far as we know there is no theoretical explanation of that phenomenon so far and we are looking forward to new experiments helping to clarify this important question.

%%%%%%%%%%%%%%%%%%%%%%%%%%%%%%%%%%%%%%%%%%%%%%%%%%%%%%%%%%%%%%%%%%%%%%

\section*{Acknowledgments}

We would like to thank Tom Bielefeld, Rolf D. Henkel, Jens Otterpohl, Klaus Pawelzik, Roland Rothenstein, Peter Ryder, Heinz Georg Schuster and Helmut Schwegler for fruitful discussions.

%%%%%%%%%%%%%%%%%%%%%%%%%%%%%%%%%%%%%%%%%%%%%%%%%%%%%%%%%%%%%%%%%%%%%%

\end{document}